\documentstyle[aps,prl]{revtex}

\newcommand{\II}{{\rm i}}

\begin{document}

\input epsf.sty
\twocolumn[\hsize\textwidth\columnwidth\hsize\csname %
@twocolumnfalse\endcsname

\draft

\widetext
\title{From the Dynamically Nanostructured Liquid to the Glassy State:
A Phenomenological Approach}
\author{J. K. Kr\"{u}ger,$^a$ J. Petersson,$^a$ J. Baller$^a$ and
M. Henkel$^b$}
\address{$^a$Laboratoire Europ\'{e}en de Recherche Universitaire
Saarland-Lorraine (LERUSL) and \\
Universit\"{a}t des Saarlandes, Fakult\"{a}t f\"{u}r Physik$^{*}$ und
Elektrotechnik 7.2, Geb\"{a}ude 38, D-66041 Saarbr\"{u}cken, Germany\\
$^b$ LERUSL et
Laboratoire de Physique des Mat\'{e}riaux,$^{**}$ Universit\'{e} Henri
Poincar\'{e} Nancy I, \\
B.P.239, F-54506 Vand{$\oe$}uvre-l\`{e}s-Nancy Cedex, France}
\date{\today}
\maketitle
\begin{abstract}
Based on experimental evidences we present a phenomenological
description of the thermal glass transition as a dynamical phase
transition. Different susceptibilities calculated on the basis of
this description are in good qualitative agreement with
experimental data. As an essential feature this novel view of the
glass transition reflects the kinetic as well as the transition
aspect of the thermal glass transition.
\end{abstract}

\phantom{.}
]

\narrowtext

In the last twenty years of theoretical and experimental work on
glasses a tremendous amount of details has been elucidated, see
\cite{ANG00,DON92,DEB96,ELL90,GUT95}. The most impressing
advances in the understanding of the freezing of liquids were
obtained by the mode coupling theory (MCT) \cite{GOE99} and
through numerical simulations \cite{BIN99}. Unfortunately, there
seems to be no direct bridge \cite{DEB01} from the MCT transition
to the thermal glass transition (TGT) because the first one
occurs at about 40~K above the TGT. The slow progress in the
understanding of the glass problem is probably due to the
contradictory experimental findings: The TGT shows clear kinetic
as well as phase transition features
\cite{JKK96CMN,JKK86CPS,JKK99PRB,JKK99NM}. Because of the obvious
kinetic features the majority of the publications in the field
(see e.g. \cite{DON92,JAE86,BIN96}) follows the idea that the TGT
is due to an accidental cross-over of time scales, i.e. a
cross-over of the structural relaxation times (of the
{$\alpha$}-relaxation) with the experimental time scales
involved. There exist however serious theoretical \cite{GIB58}
and experimental \cite{JKK96CMN,JKK86CPS,JKK99PRB,JKK99NM} works
which clearly indicate the existence of a well defined glass
transition temperature $T_{gs}$. Our own studies, performed with
a novel time domain Brillouin light scattering technique led us
to the conclusion that an intrinsic glass transition indeed
exists. Moreover the crossing of time scales can be overcome and
consequently the intrinsic glass transition at $T_{gs}$ can be
detected experimentally \cite{JKK96CMN,JKK86CPS}.

Based on experimental facts, we present the following
phenomenological description of the TGT as a dynamic phase
transition which merges both the influences of the experimentally
confirmed transition- and kinetic influences due to the thermal
history of the sample on those susceptibilites which are affected
by the TGT. We shall test our phenomenological approach with
respect to some important experimental facts, which any reliable
model calculation should reflect:

i.) The (operative) thermal glass transition temperature, $T_{g}$,
is displayed  by kink-like, jump-like or bending anomalies in static
quantities like the mass density, the elastic stiffness
constants, the enthalpy, the volume expansion coefficient etc.
\cite{DON92,JAE86}.
The static shear stiffness $c_{44}^{s}(T)$ jumps at the glass
transition temperature $T_{gs}$ yielding $c_{44}^{s}(T)=0$ for
$T>T_{gs}$ and $c_{44}^{s}(T)=c_{44}^{glass}(T)$ for $T \leq
T_{gs}$. ii.) A dynamic glass transition temperature $T_{g}^{dyn}
\geq T_{gs}$ can be defined by the cross-over of the probe
frequency $\omega$ with the structural relaxation frequency
$\omega_{\alpha}$ which accompanies the freezing process.
The structural $\alpha$-relaxation time $\tau_{\alpha} =
\omega_{\alpha}^{-1}$ is proportional to the shear viscosity of
the glass-forming material. At probe frequencies in the GHz
regime, the difference $T_{g}^{dyn} - T_{gs}$ usually exceeds 100
K. The temperature interval $T_{gs} \leq T \leq T_{g}^{dyn}$
for which $\omega \tau_{\alpha} \gg 1$ holds is called the slow
motion regime. Susceptibilities determined in the slow motion
regime are dynamically frozen quantities. In spite of their
dynamically clamped character, these susceptibilities like the
hypersonic shear stiffness or the refraction index show a
kink-like anomaly \cite{MEI86} at the {\em same}
temperature as static quantities like the specific volume
\cite{KOV58,SCH80,REH80}. The only possibility to relate this
kink anomaly to a cross-over of time scales is to find another
experimental time-scale independent of the probe frequency. The
remaining time usually available in an experiment is due to the
cooling scenario. iii.) These kink-like anomalies sharpen with
decreasing cooling rates and cannot be shifted continuously to
lower temperatures by a further decrease of the cooling rates,
there seems to exist a lower boundary for $T_{g}$
yielding $Min(T_{g})=T_{gs}$
\cite{JKK96CMN,JKK86CPS,JKK99PRB,JKK99NM}. iv.) $T_{g}$ is
usually derived from the intersection point of the high
temperature with the low temperature asymptote in the case of
kink-like and bending anomalies, or by inflection- or
onset-points in the case of jump-like anomalies
\cite{DON92,JAE86}. v.) $T_{g}$ depends on the heating or cooling
rate and thus shows kinetic features \cite{DON92,JAE86}. vi.)
Although the absolute values of phenomenological susceptibilities
within the glassy state may depend on the thermal history of the
sample, {\it the temperature derivatives of these
susceptibilities remain almost invariant} \cite{DON92,JAE86}.
vii.) From the structural point of view the TGT is an
isostructural transition. viii.) Usually the structural glass
relaxations ({$\alpha$}-process) follow a Vogel-Fulcher-Tamman
(VFT) behaviour. {\it However, on an extremely long time scale,
close to the intrinsic glass transition, an Arrhenius law with a
cut-off frequency was observed, see fig. \ref{fig_activation}}
\cite{JKK96CMN,JKK86CPS}. ix.) Neither the limiting temperature
$T_{0}$ due to the VFT behavior nor the Kauzmann temperature
$T_{K}$ of the entropy catastrophe has been observed
experimentally \cite{JKK96CMN,JKK86CPS,JKK99PRB,JKK99NM}. x.) The
elastic anharmonicity displayed by the longitudinal mode
Gr\"{u}neisen parameter behaves discontinuously at the thermal
glass transition and this discontinuity cannot be removed by
annealing \cite{JKK96JPCM,JKK96PRB}.

The basis for the following phenomenological description is the
cut-off behaviour of $\tau_{\alpha}$ shown in fig.
\ref{fig_activation} \cite{JKK96CMN,JKK86CPS,JKK99PRB,JKK99NM}
\begin{equation}
\tau_{\alpha}(T)=\left\{
\begin{array}{ll}
A\, {\rm e}^{{{\Delta}G}/{RT}} & ~;~~ T \geq T_{gs}\\
\infty & ~;~~T < T_{gs}
\end{array}
\right. \label{tau_alpha}
\end{equation}
with $A$: inverse attempt frequency, $\Delta G$: activation
enthalpy, $R$: universal gas constant. A similar Arrhenius law
with a clear cut-off at a distinguished temperature was found by
L\"{u}ty et {\em al.} for the order-disorder transition of
CN\raisebox{1ex}{-} molecules \cite{LUE83}. We assume as a
working hypothesis that the TGT is a dynamic phase transition
which is governed by the temperature dependent relaxation time
given by eq. (\ref{tau_alpha}). The model is moreover based on
the fact that the TGT shows besides its phase transition
character (i, ii, viii, x) obvious kinetic features including
non-ergodicity (iii, iv, v, vi, ix). Thus it can hardly be
described in the setting
of a classical order parameter scenario and critical
slowing-down. We assume that within the high-temperature phase
the freezing process is accompanied by the formation of
nano-sized regions of which we think as of dynamic randomly
closed packed clusters (rcp's). Sufficiently close but above
$T_{gs}$ the rcp's are dynamically percolated in the sense that
they form a temporary chain-network consisting of rcp's which
have an average life-time $\tau_{\alpha}(T)$. $\tau_{\alpha}(T)$
may be viewed as a temperature-dependent internal clock within the
material which defines a unique intrinsic glass transition
temperature $T_{gs}$ by the condition that $\omega_{\alpha} =
\tau_{\alpha}^{-1}(T)$ drives all dynamic properties relevant for
the freezing process. Thus we admit that the transition parameter
$\omega_{\alpha}(T)$ controls the glass transition simultaneously
in three areas: a) Influence on the structure (see below). b)
Dynamic influence: response to a periodic perturbation (probe
frequency). c) Kinetic influence: response to temperature
perturbation (time domain).

The structural properties close to the the TGT are determined by
the mass ratio $x^{rcp}=m^{rcp}/m$ where $m$ is the total mass of
the sample and $m^{rcp}$ the total mass of rcp's.
If every relevant experimental time constant
$\tau_{exp}$ is always larger than the instrinsic time constant
$\tau_{\alpha}(T)$,
we assume that the concentration of the
rcp's is zero at sufficiently high temperatures and has to go
continuously to one on approaching $T_{gs}$. Taking into account
that $x^{rcp}(T)$ is intimately related to the specific volume of
the glass-forming material, its temperature derivative must
jump at $T_{gs}$ although this step in the volume
expansion coefficient is often smeared out in experimental data.
These basic requirements can be fulfilled  by a simple relation depending
only on the transition parameter
\begin{equation}
x^{rcp}(T)=\left\{
\begin{array}{ll}
{\rm e}^{-(\omega_{\alpha}(T)-\omega_{\alpha}(T_{gs}))} & ~;~~ T \geq T_{gs}\\
1 & ~;~~ T < T_{gs}
\end{array}
\right. \label{x_rcp}
\end{equation}
Eq. (\ref{x_rcp}) incorporates the idea that at temperatures
$T_{gs}+\epsilon$ (where $\epsilon\ll 1$),
a stationary interchange between liquid and rcp-phase takes place.
As soon as the minimum frequency $\omega_{\alpha}(T_{gs})$ is
reached, any molecular dynamic related to the liquid-glass
transition becomes so slow that it spontaneously dies.
Although (\ref{x_rcp}) is an {\it ad hoc} assumption, it works
remarkably well in model calculations.

According to our model and in line with experimental data the
transition parameter $\omega_{\alpha}(T)$ furthermore controls
dynamic properties like the shear stiffness $c_{44}(T)$ or the
shear viscosity $\eta_{44}(T)$. In order to demonstrate the
effect of $\omega_{\alpha}(T)$ on $c_{44}(T)$ in the vicinity of
the glass transition we use
the Debye-process as the simplest relaxation model.
Other relaxation models can easily be incorporated in our
phenomenological approach. The real part of the dynamic shear
stiffness $c_{44}^{*}(\omega)$ is given by
\begin{equation}
c_{44_{Debye}}^{*}(\omega) = \II \omega \eta_{44}^{*} = \frac{\II
\omega \tau_{\alpha} c_{44}^{\infty}}{1+ \II \omega \tau_{\alpha}}
\label{c44_debye}
\end{equation}
If the experimental time constant $\tau_{exp_{T}}$ of the cooling
or heating process becomes equal or smaller than the structural
relaxation time $\tau_{\alpha}$, the sample will fall out of
thermal equilibrium. In other words, in the vicinity of the glass
transition the phonon bath controlling the temperature of the sample is
already in equilibrium whereas the dynamic ``nano-composit'' is
not. Any quantity $Q(T)$ describing thermal recovery processes
close to $T_{gs}$ is expected to depend on $\omega_{\alpha}(T)$.
A simple expression which takes into account the behaviour close
to $T_{gs}$ and the dying-out for high $T$ and large
$\tau_{exp_{T}}$ is
\begin{equation}
Q(T) = 1-q_0 \left\{
\begin{array}{ll}
{\rm e}^{-{{\tau_{exp_{T}}}/{\tau_{\alpha}(T)}}} & ~;~~ T \geq T_{gs}\\
{\rm e}^{-{{\tau_{exp_{T}}}/{\tau_{\alpha}(T_{gs})}}} & ~;~~ T <
T_{gs}
\end{array} \right. \label{Q_T}
\end{equation}
Here $q_0$ describes
the degree of possible quenching.
Even in the long-time limit the permanent creation and
annihilation of rcp's within the glass-forming liquid permanently
changes the network structure and thus hinders the formation of a
rigid network above $T_{gs}$.
The appearance of the glass transition will strongly depend on
the relative magnitude of each individual influence given by
(\ref{x_rcp}), (\ref{c44_debye}) and (\ref{Q_T}).

Experimentally, the specific volume in the glassy
state as well as in the liquid state behaves linear over a wide
range of temperatures (e.g. \cite{KOV58,SCH80,REH80}) but the
volume expansion coefficients of the liquid state are higher than
those of the glassy state ($\alpha_l > \alpha_g$). The position of
the operative $T_g$ depends on the cooling rate whereas $\alpha_l$
and $\alpha_g$ remain unchanged \cite{REH80}. If
the mass density is measured with a purely static experimental
technique, the only relaxation phenomenon expected is that which
emerges from the temperature perturbation on approaching $T_{gs}$,
being expressed by the experimental time scale $\tau_{exp_T}$. We
assumed above
that sufficiently close but
above $T_{gs}$ the rcp's are above the percolation threshold. This
condition guarantees the onset of mechanical rigidity of
glass-forming samples at $T_{gs}$ (i).
When approaching $T_{gs}$ from above, from (\ref{x_rcp}) the rcp concentration
increases towards one.
At the same time the related network needs increasing time in
order to reach its dynamic equilibrium structure.
The equilibrium specific volume $v^{eq}(T)$ depends on the
specific volume of the liquid part $v^{l}(T) = v^{l_0}(1+\alpha^l
(T-T_{gs}))$, on the rcp part
$v^{rcp}(T) = v^{rcp_0}(1+\alpha^{rcp} (T-T_{gs}))$, and on $x^{rcp}(T)$
\begin{equation}
v^{eq}(T) = v^l(T)(1-x^{rcp}(T))+v^{rcp}(T) x^{rcp}(T)
\label{v_eq}
\end{equation}
Thus the ideal glass transition described by $v^{eq}(T)$ is
determined by the kink-like anomaly of $x^{rcp}(T)$ at $T_{gs}$.
In order to incorporate the competition between the thermal
treatment given by $\tau_{exp_T}$ and the
transition parameter $\omega_{\alpha}$ on the specific volume given by
eq. (\ref{v_eq}), we use the following simple relation
\begin{equation}
v(T) = Q(T,\tau_{exp_T}) v^{eq}(T)  \label{v_T}
\end{equation}

We have calculated the relaxing specific volume of a rcp-liquid
composite above and below the glass transition using the
relaxation data obtained for polyvinylacetate (PVAc)
\cite{JKK96CMN,JKK86CPS,JKK99PRB,JKK99NM}. For $v^l(T)$ and
$v^{rcp}(T)$ we have anticipated the linear temperature
dependencies given above. Fig. \ref{fig_model} gives the
calculated specific volume curves for different experimental time
constants. These results are in reasonable qualitative agreement
with specific volume data \cite{KOV58}. In
accordance with experimental observations for increasing cooling
rates the $v(T)$-curves become increasingly bent and the operative
transition anomaly expected at $T_g$ is gradually shifted to
higher temperatures with decreasing experimental time constants
$\tau_{exp_T}$. Moreover, with increasing cooling rates the true
transition at $T_{gs}$ becomes more and more hidden.
The operative glass transition temperatures $T_g$ shown in fig.
\ref{fig_model} are unrelated to the true transition at
$T_{gs}$ but are a measure for the cross-over of the
transition frequency $\omega_{\alpha}$ with the experimental time
scale $\tau_{exp_T}$.

At high probe frequencies, the jump-like discontinuity of the
static shear stiffness at $T_{gs}$ transforms into a kink-anomaly
(ii), which at even higher temperatures (transition from the slow
motion to the fast motion regime) is followed by a mechanical
slowing-down to zero.
If a glass-forming liquid relaxes according to a Debye law,
one calculates from (\ref{c44_debye}) the real part
of the complex dynamic shear modulus of a liquid \cite{HER59} as
\begin{equation}
c_{44_{Debye}}^{'}(T, x_{rcp}(T),
\tau_{exp_{\omega}})=\frac{c_{44}^{\infty}(T,
x^{rcp}(T))}{1+(\omega_{\alpha}(T) \tau_{exp_{\omega}})^2}
\label{c44_real}
\end{equation}
where $c_{44}^{\infty}(T, x^{rcp}(T))$ is the temperature-dependent
shear stiffness as measured at very high frequencies
which close to but still above TGT may significantly depend on
the thermal treatment of the sample. By our definition,
$c_{44}^{\infty}(T, x^{rcp}(T))$ depends on the concentration of
the rcp's and on the temperature. In order to account for the
effect of $x^{rcp}(T)$ on $c_{44}^{\infty}(T, x^{rcp}(T))$ we use
the Voigt model \cite{SEN88}
to account for the elastic interaction of the liquid part
$c_{44}^l(T)$ and the rcp part $c_{44}^{rcp}(T)$ within the sample
\begin{displaymath}
c_{44}^{\infty}(T, x^{rcp}(T)) = c_{44}^l(T)
(1-x^{rcp}(T))+c_{44}^{rcp}(T) x^{rcp}(T)  \label{voigt}
\end{displaymath}
We assume
for the elastic modulus of the liquid part
$c_{44}^{l}(T) = c_{44}^{l_0} (1 + \alpha_{44}^{l} (T-T_{gs}))$
and for the rcp part
$c_{44}^{rcp}(T) = c_{44}^{rcp_0} (1 +\alpha_{44}^{rcp}
(T-T_{gs}))$ where $c_{44}^{l_0}, \alpha_{44}^l, c_{44}^{rcp_0},
\alpha_{44}^{rcp}, T_{gs}$ are constants.
Since the mechanical properties of the rcp's are expected to
deviate only slightly from those of the liquid phase, the measured
data will be rather insensitive with respect to the chosen
composite model (Voigt or Reuss)\cite{SEN88}. Finally we
take into account the effect of the cooling/heating procedure on the
elastic shear stiffness
by combining eqs. (\ref{Q_T},\ref{c44_real}).
It turns out that the influence of the
thermal history on the recovery of the elastic constants can be
incorporated in the same formal way as it was done for
the specific volume, yielding
\begin{equation}
c_{44}^{'}(T) = Q(T, \tau_{exp_T}) c_{44_{Debye}}^{'}(T)
\label{c44_final}
\end{equation}
According to the results shown in fig. \ref{fig_model}, eq.
(\ref{c44_final}) describes at least qualitatively the static and
dynamic elastic behavior around the glass transition including
thermal recovery effects. The genuine discontinuous change of the
static shear modul at $T_{gs}$ is well described. As mentioned
above we have for the first time observed the discontinuous
behaviour of the $\alpha$-relaxation process for PVAc
\cite{JKK96CMN,JKK86CPS,JKK99PRB,JKK99NM}.
Therefore the $c_{44}^{'}(T)$-curves in fig. \ref{fig_model} have
been calculated on the basis of data for the activation energy
and the attempt frequency of the $\alpha$-process of this
material. Since PVAc has an
almost vanishing scattering cross-section for shear phonons,
the shear modulus is hard to measure.
However the qualitative behaviour of the dynamic shear modulus
shown in fig. \ref{fig_model} was often reported in literature
(e.g. \cite{JKK80,JKK84}). The cross-over of the hypersonic probe
frequency with the internal glass relaxation process happens at
temperatures well above $T_{gs}$. This indicates
that the kink-like anomaly close to $T_{gs}$ has {\em nothing}
to do with a cross-over of the probe frequency with the
$\alpha$-relaxation time.

In conclusion, we have introduced a new working hypothesis
treating the ideal glass transition as a dynamic phase
transition. We identify the jump in the transition parameter
$\omega_{\alpha}(T)=\tau_{\alpha}^{-1}(T)$ at $T=T_{gs}$, see eq.
(\ref{tau_alpha}), as the fundamental feature of the glass
transition. This sole feature, when combined with standard
knowledge on glasses, allows one to predict the temperature- and
time- dependence of susceptibilities sensitive to the glass
transition in good agreement with experimental data, reflecting
at the same time phase transition and kinetic phenomena.


\begin{figure}
\caption{Activation Plot of
$\omega_{\alpha}(T)=\tau_{\alpha}^{-1}(T)$ taken from refs.
\protect\cite{JKK96CMN,JKK86CPS,JKK99PRB,JKK99NM}.
At temperatures below (a) no further relaxation could be
found (see text). \label{fig_activation}}
\end{figure}
\begin{figure}
\caption{Specific volume $v$ (curves a-c) and shear stiffness
$c_{44}$ (curves d-f) as a function of temperature calculated
from eqs. (\ref{v_T},\ref{c44_final}) with different experimental
time constants: (a) $\tau_{exp_{T}} = 10^9 s$  (b)
$\tau_{exp_{T}} = 10^3 s$ (c) $\tau_{exp_{T}} = 10^{-1} s$ and
(d) $\tau_{exp_{T}} = 10^8 s$, $\tau_{exp_{\omega}} = 10^8 s$ (e)
$\tau_{exp_{T}} = 10^2 s$, $\tau_{exp_{\omega}} = 10^{-4} s$ (f)
$\tau_{exp_{T}} = 10^{-1} s$, $\tau_{exp_{\omega}} = 10^{-6} s$.
$T_{g1}, T_{g2}$: operative glass transition temperatures.}
\label{fig_model}
\end{figure}

\newpage
\widetext

\begin{figure}
\centerline{\epsfxsize=7.2in\epsfbox{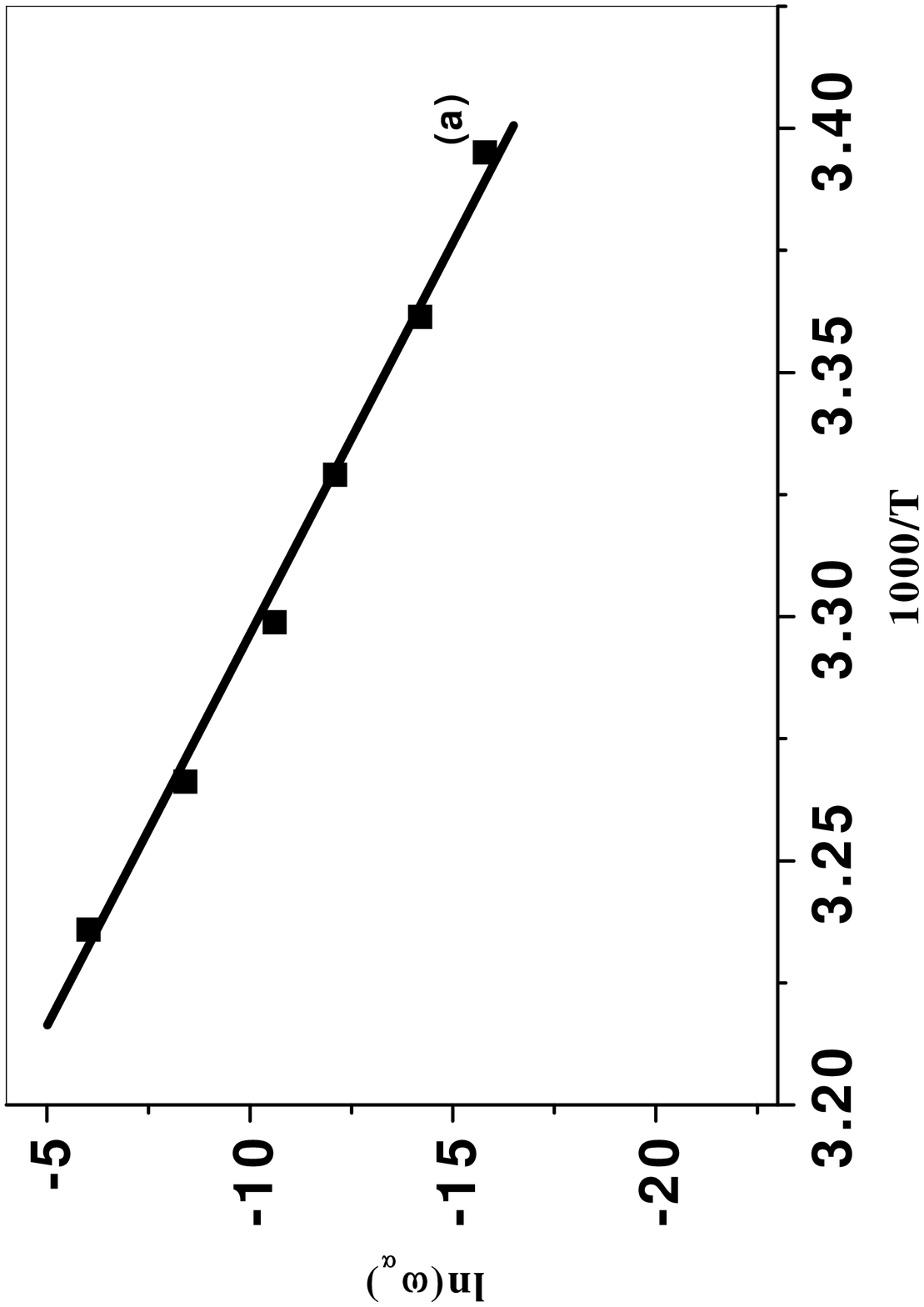}}
\end{figure}
~
\newpage

\begin{figure}
\centerline{\epsfxsize=7.2in\epsfbox {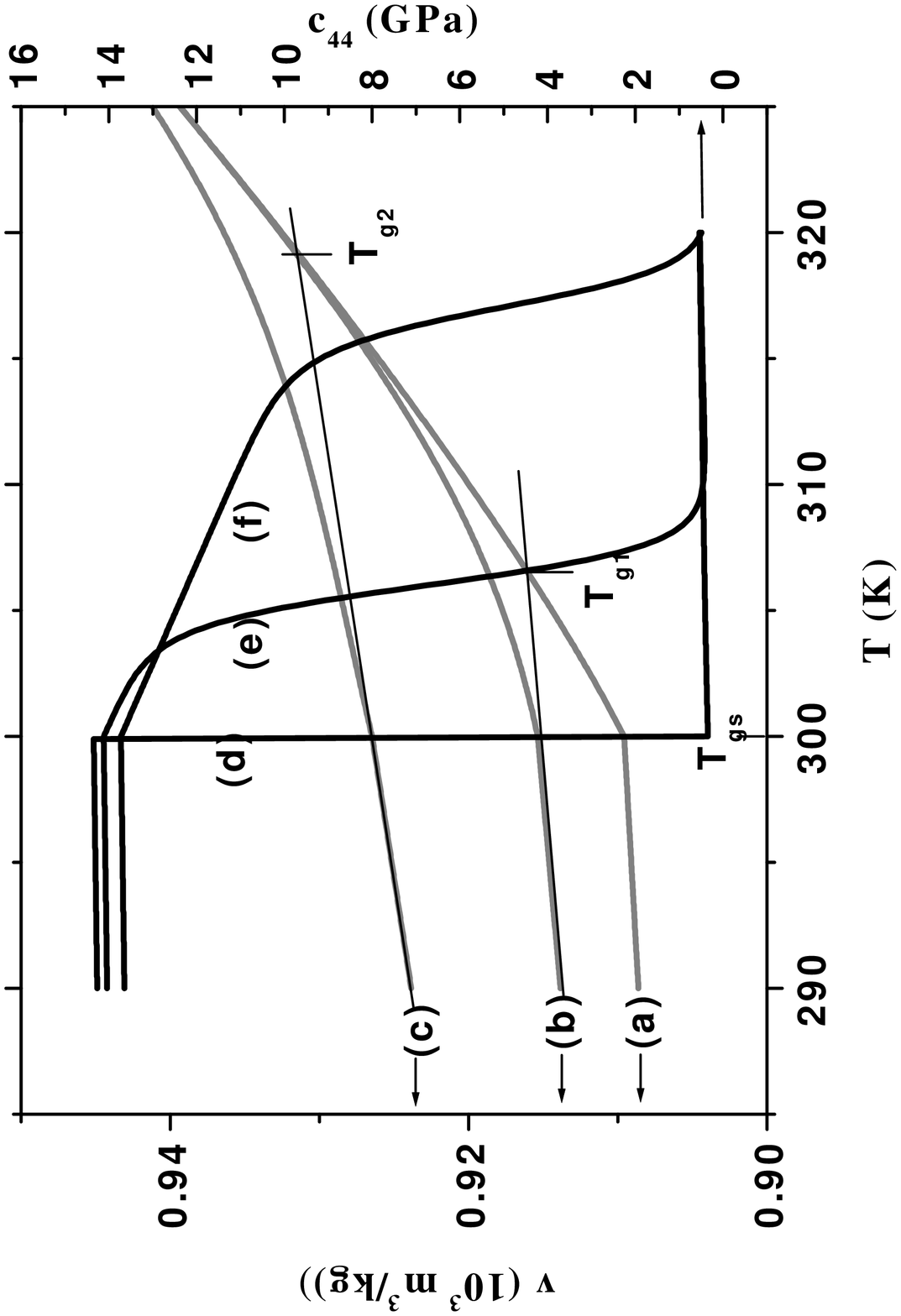}}
\end{figure}

\end{document}